\documentstyle [12pt]{article}
\def\be{\begin{equation}}
\def\ee{\end{equation}}
\def\bea{\begin{eqnarray}}
\def\eea{\end{eqnarray}}
\def\pd{\partial}

\begin{document}

\begin{titlepage}

\title{The Complex Bateman  Equation}

\author{D.B. Fairlie$\footnote{e-mail: david.fairlie@durham.ac.uk}$\\
\quad\\
{\it Department of Mathematical Sciences}\\
{\it University of Durham, Durham DH1 3LE}\\
and\\
 A.N. Leznov$\footnote{e-mail: leznov@mx.ihep.su. }$\\
\quad \\
{\it  Institute for High Energy Physics, 142284 Protvino,}\\{\it Moscow Region,
Russia}\\
{\it and}\\ 
{\it  Bogoliubov Laboratory of Theoretical Physics, JINR,}\\
{\it 141980 Dubna, Moscow Region, Russia}}
\maketitle

\begin{abstract}
The general solution to the Complex Bateman equation is constructed. It is given in implicit form in terms of a functional relationship for the unknown function.
The known solution of the usual Bateman equation is recovered as a special case.
\end{abstract}
\end{titlepage}
\section{Introduction}
 We define the Complex Bateman equation as
\be 
\det\left|\begin{array}{ccc}0&\frac{\pd \phi}{\pd y_1}&\frac{\pd \phi}
{\pd y_2}\\
\frac{\pd \phi}{\pd x_1}&\frac{\pd^2\phi}{\pd x_1\pd y_1}&\frac{\pd^2\phi}{\pd x_1\pd y_2}\\
\frac{\pd \phi}{\pd x_2}&\frac{\pd^2\phi}{\pd x_2\pd y_1}&\frac{\pd^2\phi}{\pd x_2\pd y_2}\\
\end{array}\right|\,=\,0.\label{batman}
\ee
The real form of the Bateman equation;
\be
(\frac{\pd\psi}{\pd x})^2\frac{\pd^2\psi}{\pd y^2}+(\frac{\pd\psi}{\pd y})^2\frac{\pd^2\psi}{\pd x^2}-2\frac{\pd\psi}{\pd x}\frac{\pd\psi}{\pd y}\frac{\pd^2\psi}{\pd x\pd y}\,=\,0.\label{realbat}
\ee
arises when the unknown function depends upon only two arguments,
$x\,=\,x_1+y_1,\ y\,=\,x_2+y_2$

This equation, which is equivalent to the pair of first order equations
\bea 
\frac{\pd u}{\pd x}&=&u\frac{\pd u}{\pd y}\nonumber\\
\frac{\pd \psi}{\pd x}&=&u\frac{\pd \psi}{\pd y},\label{first}
\eea
has a general solution given implicitly by choosing two arbitrary functions of
one variable $f(\psi)$ and $g(\psi)$ and constraining them to satisfy the linear relation
\be
xf(\psi)\,+\,yg(\psi)\,=\, c\,=\,\quad{\rm constant}.\label{constraint}
\ee
An inhomogeneous form of the Bateman equation, the so-called two dimensional Born-Infeld equation, is equivalent to the equation describing minimal surfaces,
and has been solved by Bateman himself, \cite{bat} and Barbasov and Chernikov \cite{bar}
Further properties and generalisations of the real Bateman equation
can be found in \cite{fai}\cite{fab}.
The general solution to the complex Bateman equation, (\ref{batman}), on the other hand, is given, again implicitly, by identifying two arbitrary functions of three variables, $F(\phi,x_1,x_2)$ which depends upon $(\phi,\ x_1,\ x_2)$ and $G(\phi,y_1,y_2)$ depending upon $(\phi,\ y_1,\ y_2)$ and solving the resulting equality 
\be
F(\phi,x_1,x_2)\,\equiv\,G(\phi,y_1,y_2).\label{claim} 
\ee
implicitly for $\phi(x_1,x_2,y_1,y_2)$.   This assertion may be readily verified. This solution encompasses the above solution to the real Bateman equation  by the choice of the arbitrary functions $F,\ G$ as
\be
F\,=\,x_1f(\phi)+y_1g(\phi),\quad\quad G\,=\,-x_2f(\phi)-y_2g(\phi) +c.
\label{choice}
\ee

 It is the purpose of this note to explain  how this  result may be deduced, with an eye to further generalisation. At this point we insert a caveat; this analysis is carried out in the spirit of many investigations in mathematical physics, of being a little cavalier about rigorous questions of differentiability of the functions involved. We assume that the functions
with which we deal are twice differentiable, though we are well aware that the real Bateman equation admits solutions of shock wave type, where differentiability at one or more points fails.

\section{Proof}
The complex Bateman equation, (\ref{batman}) is the eliminant of three linearly
dependent equations which may be written as;
\bea \alpha^1\frac{\pd\phi}{\pd y_1}\ +\ \alpha^2\frac{\pd\phi}{\pd y_2}&=&0\nonumber\\
\frac{\pd\phi}{\pd x_1}-\frac{\pd\alpha^1}{\pd x_1}\frac{\pd\phi}{\pd y_1}-\frac{\pd\alpha^2}{\pd x_1}\frac{\pd\phi}{\pd y_2}&=&0\label{elim}\\
\frac{\pd\phi}{\pd x_2}-\frac{\pd\alpha^1}{\pd x_2}\frac{\pd\phi}{\pd y_1}-\frac{\pd\alpha^2}{\pd x_2}\frac{\pd\phi}{\pd y_2}&=&0\nonumber
\eea
Here $\alpha^1,\ \alpha^2$ are functions of the variables $(x_1,\ x_2,\ y_1,\
y_2)$.
 The linear equations whose eliminant gives  (\ref{batman}) are obtained 
from the first equation of (\ref{elim}) together with the second plus the derivative of the first with respect to $x_1$ and the third plus the $x_2$ derivative of the first.  These equations admit an obvious generalisation to any number of pairs $x_i,y_j$.
 Clearly, 
\bea 
\frac{\pd \phi}{\pd x_1}-\frac{\pd}{\pd x_1}\left(\log(\alpha^1)-\log(\alpha^2)\right)\frac{\pd\phi}{\pd y_2}\alpha^2&=&0\label{one}\\
\frac{\pd \phi}{\pd x_2}-\frac{\pd}{\pd x_2}\left(\log(\alpha^1)-\log(\alpha^2)\right)\frac{\pd\phi}{\pd y_2}\alpha^2&=&0.\label{two}
\eea
Cross differentiation shows that $(\frac{\pd\phi}{\pd y_2}\alpha^2)^{-1}$ is a function of $(\phi,\ y_1,\ y_2)$ and hence we may write
\be 
\log\left(\frac{\alpha^1}{\alpha^2}\right)\,=\, K(\phi,y_1,y_2),\label{resolve}
\ee
where $K$ is an arbitrary function of  $(\phi,\ y_1,\ y_2)$. Further implications of these equations are as follows;
\be
\frac{\pd\phi}{\pd y_1}\,=\,-\frac{1}{K'\alpha^1},\quad\frac{\pd\phi}{\pd y_2}\,=\,\frac{1}{K'\alpha^2},\label{imply2}
\ee 
where $K'$ denotes the partial derivative of $K$ with respect to $\phi$. As a consequence, $\exists$
 a function $U(\phi,y_1,y_2)$ such that
\be
\frac{\pd\phi}{\pd y_1}\,=\,U(\phi,y_1,y_2)\frac{\pd\phi}{\pd y_2}
\label{exist1}
\ee
Similarly, we can introduce a second function $V(\phi,x_1,x_2)$ such that
\be
\frac{\pd\phi}{\pd x_1}\,=\,V(\phi,x_1,x_2)\frac{\pd\phi}{\pd x_2}
\label{exist2}
\ee
The integrability condition for those two equations, obtained by eliminating the mixed $(x_1,\ y_1)$ derivatives of $\phi$ is automatically satisfied. If $U(\phi,y_1,y_2)$  is written in the form
\be
U\,=\, \frac{\frac{\pd}{\pd y_1} G(\phi,y_1,y_2)}{\frac{\pd}{\pd y_2} G(\phi,y_1,y_2)}\label{write}
\ee
for some function $G(\phi,y_1,y_2)$,
where $\phi$ is regarded as a parameter the partial derivatives with respect to $y_1$ and $y_2$ act on the last two arguments of  $G$ , then this is  simply a first order differential equation for $G$, which is in principle solvable.
However the partial derivatives in (\ref{write}) may be replaced by total derivatives  when $\phi$ is now regarded as a function of $(x_1,\ x_2,\ y_1,\ y_2)$  since
 \be
U\,=\, \frac{\frac{dG}{d y_1}}{\frac{dG}{d y_2}}= 
 \frac{\frac{\pd}{\pd \phi} G(\phi,y_1,y_2)\frac{\pd\phi}{\pd y_1}+\frac{\pd}{\pd y_1} G(\phi,y_1,y_2)}{\frac{\pd}{\pd \phi} G(\phi,y_1,y_2)\frac{\pd\phi}{\pd y_2}+\frac{\pd}{\pd y_2} G(\phi,y_1,y_2)}=
\frac{\frac{\pd\phi}{\pd y_1}}{\frac{\pd\phi}{\pd y_2}}
\label{write2}
\ee
using (\ref{exist1}). This equation 
 implies that
$G$ is a function of $\phi$, together with the additional variables in the problem, i.e. $x_1,\ x_2$. So we may write
\be
 G(\phi,y_1,y_2)\,=\,F(\phi,x_1,x_2)
\label{proof}
\ee
for some function $F$, which is the result announced.

 By the same token,
\be
V\,=\, \frac{\frac{d}{d x_1} F(\phi,x_1,x_2)}{\frac{d}{d x_2} F(\phi,x_1,x_2)}\label{write1}
\ee

\section{Conclusions}  We have shown that what we have called the complex Bateman equation,
\be
\frac{\pd\phi}{\pd x_1}\frac{\pd\phi}{\pd y_1}\frac{\pd^2\phi}{\pd x_2\pd y_2}+\frac{\pd\phi}{\pd x_2}\frac{\pd\phi}{\pd y_2}\frac{\pd^2\phi}{\pd x_1\pd y_1}-\frac{\pd\phi}{\pd x_1}\frac{\pd\phi}{\pd y_2}\frac{\pd^2\phi}{\pd x_2\pd y_1}-\frac{\pd\phi}{\pd x_2}\frac{\pd\phi}{\pd y_1}\frac{\pd^2\phi}{\pd x_1\pd y_2}\,=\,0.\nonumber
\ee
may be solved completely in terms of two arbitrary functions $F(\phi,y_1,y_2)$ and $G(\phi,x_1,x_2)$  which are constrained to be equal.
The first order equations (\ref{exist1}) and (\ref{exist2}) both separately
imply the complex Bateman equation.
We expect that the extension of these results to higher dimensions will proceed along similar lines to that for the real Bateman equation. \cite{lez1}\cite{lez2}. 
 We hope to return to the question of the solution of the complex generalisation in arbitrary dimensions in the near future.

\end{document}